# Targeted-Subharmonic-Eliminating Pulse Density Modulation for Wireless Power Transfer Systems

Songyan Li, *Student Member, IEEE*, Hongchang Li, *Senior Member, IEEE*, Haiyue Jiang, *Student Member, IEEE*, Yudong Zhang, *Student Member, IEEE*, Wenjie Chen, *Senior Member, IEEE*, Xu Yang, *Senior Member, IEEE*,

*Abstract*—This letter proposes a targeted-subharmonic-eliminating pulse density modulation (PDM) method for series-series (SS) compensated wireless power transfer (WPT) systems. The subharmonic frequency components which excite current abnormal oscillations in PDM controlled WPT systems are eliminated through a specially designed noise transfer function (NTF). The proposed method is simple to implement in both primary and secondary sides of WPT systems and exhibits a certain tolerance to deviations caused by inaccurate coupling coefficient identification in NTF design. Experimental results demonstrated the effectiveness and robustness of the proposed method in suppressing current abnormal oscillations and reducing the fluctuations in current amplitudes.

*Index Terms*—Wireless Power Transfer (WPT), pulse density modulation (PDM), noise transfer function (NTF), subharmonic, zero-voltage switching (ZVS).

## I. INTRODUCTION

Pulse density modulation (PDM) is a promising approach for high-efficiency power conversion, which has become a research hotspot in wireless power transfer (WPT) systems. Unlike phase shift modulation, which suffers from hard switching, PDM consistently maintains soft-switching operation with low reactive power and reduced average switching frequency, while achieving wide-range voltage gain adjustment in both primary and secondary sides [1]-[2].

Delta-sigma pulse density modulation (ΔΣ-PDM) enables continuous modulation with theoretically arbitrary pulse density resolution, and is ideal for efficiency-optimized closed-loop control WPT systems. The spectrum of modulated wave generated by the ΔΣ-modulator contains abundant frequency components that are non-integer multiples of the switching frequency, namely, subharmonics. Due to the frequency-selective characteristics of series-series (SS) network, the subharmonic components are effectively suppressed, and the current amplitude typically exhibits minor fluctuations [1]-[2]. However, under specific pulse density conditions, the current demonstrates abnormal oscillations exceeding 50% in amplitude [3]-[5]. The abnormal oscillations in current increase conduction losses, disrupt soft-switching operation, and may damage the capacitors and switching devices.

Thus, various solutions have been proposed to solve this problem. In [3], an active damping control by transiently introducing phase shift modulation is proposed to suppress the current oscillations, but it causes hard switching in some cycles. In [4], by adding a sending current limiter as a condition for skipping pulses, the conditional PDM method suppresses current oscillations while maintaining soft-switching operation, but the limiting current thresholds under different coupling coefficients and DC side voltage variation have not been systematically discussed. As presented in [5], the rectifier and inverter are synchronously modulated with precisely controlled pulse density and phase alignment, leveraging the transient response superposition to eliminate oscillations, but precise current envelope detection is required to identify the pulse-skipping instances, and the method can only be implemented with dual-side control.

This letter analyzes the essential reasons for the abnormal oscillations in the SS compensated WPT system caused by ΔΣ-PDM, and proposes the targeted-subharmonic-eliminating PDM (TSE-PDM) method. The proposed modulator employs a deliberately engineered noise transfer function (NTF) with notch characteristics to eliminate subharmonic components in the modulated wave that excites abnormal oscillations.

## II. ANALYSIS OF CURRENT OSCILLATIONS

The typical circuit diagram of the SS-compensated WPT system is shown in Fig. 1, where $V_g$ ($V_o$), $U_1$ ($U_2$), $I_1$ ($I_2$), $L_1$ ($L_2$), $C_1$ ($C_2$) and $R_1$ ($R_2$) are the input (output) voltage, primary (secondary) voltage modulated wave, resonant current, inductance, capacitance, and equivalent series resistance, respectively, $M$ is the mutual inductance, and $k$ is the coupling coefficient.

The structure of a conventional ΔΣ-PDM converter implemented on the primary/secondary-side of WPT system is depicted in Fig. 2. In each half-switching cycle, the quantization error $e$ accumulates with the input pulse density $d$, and the comparator quantizes the accumulated results. The output of the quantizer $y$ is AND-ed with the pulse $c_{1/2}$ and the inverted $c_{1/2}$ to generate the driving signals for the leading-leg and lagging-leg switches, where $c_1$ is an independent input pulse when the PDM is implemented on the primary side, and $c_2$ is the synchronous pulse when the PDM is implemented on the secondary side. For the secondary-side implementation, the measured current signal is compared with zero, and the edges of the comparator output trigger the transition of the synchronous pulse $c_2$. To avoid false triggering events, a blanking window is generated after the transition of $c_2$. The driving signals of the leading-leg and lagging-leg switches are denoted by $a$ and $b$, respectively, while $a - b$ is the modulated wave, i.e. the normalized $U_{1/2}$, and $|a - b|$ is the amplitude of the modulated wave.

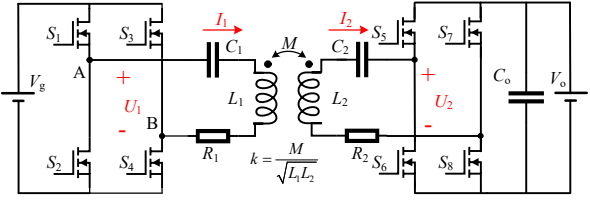

**Fig. 1.** Circuit diagram of the SS-compensated WPT system.

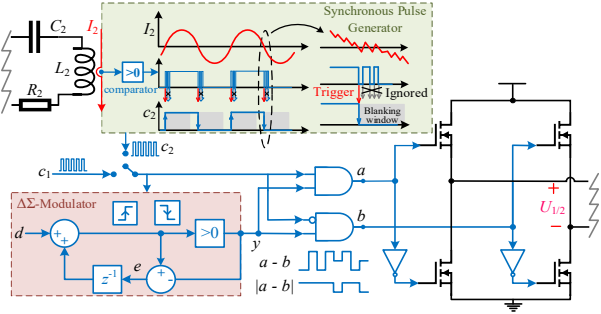

**Fig. 2.** Structure of the conventional ΔΣ-PDM converter.

As the spectrum of a density-modulated wave exhibits symmetry about the fundamental frequency $\omega_s$, a pair of subharmonics at frequencies $\omega_s \pm \Delta\omega_s$ can be equivalently represented as an amplitude-modulated fundamental wave, as:

$$\cos\left[(\omega_s + \Delta\omega_s)t\right] + \cos\left[(\omega_s - \Delta\omega_s)t\right] = 2\cos(\Delta\omega_s t)\cos(\omega_s t) \quad (1)$$

Therefore, the subharmonic excitation can be equivalently converted into fundamental-wave excitation with low-frequency amplitude modulation, and the generalized state-space averaging (GSSA) model can be employed to investigate the transfer function from voltage amplitudes to resonant current amplitudes, providing the insight of abnormal oscillations caused by specific subharmonics.

In Fig. 3, the small signal model of transfer function from the voltage amplitude to the current amplitude at $k = 0.15$ is obtained through GSSA method, with a resonant peak at $\omega_0$:

$$\omega_0 = \frac{1}{2}k\omega_s \quad (2)$$

The spectrum of the modulated wave amplitude under $d_{1/2} = 0.963$ is obtained through Fourier decomposition, as shown in Fig. 3, which contains frequency components located near the resonant peaks of the amplitude transfer functions.

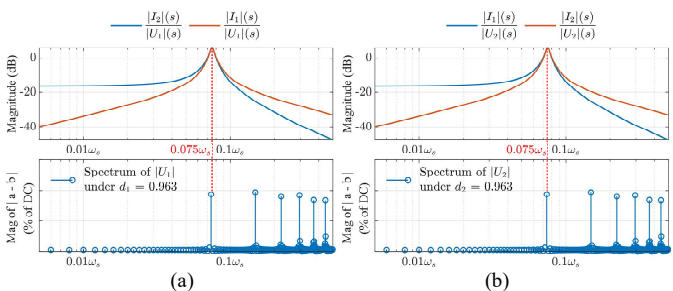

**Fig. 3.** The bode plots of voltage-to-current magnitude transfer functions at $k = 0.15$ and the spectrum of the modulated wave amplitude $|U_{1/2}|$ at $d_{1/2} = 0.963$: (a) the bode plot of $|U_1|$ to $|I_1|$ and $|I_2|$ and the spectrum of $|U_1|$, and (b) the bode plot of $|U_2|$ to $|I_1|$ and $|I_2|$ and the spectrum of $|U_2|$.

Therefore, the abnormal current oscillation caused by ΔΣ-PDM is due to the inability of SS-compensated WPT system to suppress the frequency components located at $\frac{1}{2}k\omega_s$ in the spectrum of the modulated wave amplitude $|a - b|$, or to suppress the subharmonic located at $\omega_s \pm \frac{1}{2}k\omega_s$ in the spectrum of the modulated wave $a - b$ according to (1).

## III. THE PROPOSED TSE-PDM

When the coupling coefficient $k$ is fixed, only particular subharmonics may excite oscillations. A noise shaping method to eliminate the target subharmonic is proposed to actively suppress the potential oscillation-inducing components within the modulator, thereby preventing abnormal current oscillation fundamentally.

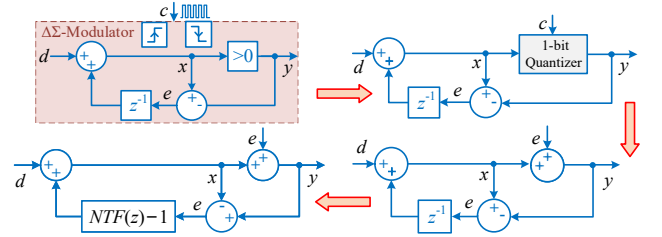

**Fig. 4.** Derivation of the linearized ΔΣ-modulator model.

The linearized model of the ΔΣ pulse-density modulator is constructed as depicted in Fig. 4. The quantization error is modeled as an additive noise term $e$, and the final modulator architecture is obtained through equivalent linear network transformation. The quantizer output $y$ equals the modulated wave amplitude $|a - b|$, which satisfies the z-domain equation:

$$Y(z) = D(z) + E(z)NTF(z) \quad (3)$$

where $Y(z)$, $D(z)$, $E(z)$ and $NTF(z)$ represent the z-transform of the quantizer output $y$, input pulse density $d$, quantization error $e$ and NTF, respectively. The NTF of the ΔΣ pulse-density modulator shown in Fig. 4 is a first-order difference block:

$$NTF_1(z) = 1 - z^{-1} \quad (4)$$

To achieve noise shaping with targeted subharmonic suppression, it is necessary to investigate a high-order NTF that differs from (4). This high-order NTF must satisfy the following four requirements:

① The DC gain of the NTF must be adequately low:

$$NTF(1) = 0 \quad (5)$$

② The NTF must be physically realizable, that is the z-domain transfer function $NTF(z) - 1$ must be zero-initialized:

$$NTF(\infty) = 0 \quad (6)$$

③ The NTF must contain a pair of complex-conjugate zeros at:

$$z_{1,2} = e^{\pm j\pi \frac{\omega_e}{\omega_s}} \quad (7)$$

where $\omega_e$ is the frequency component to be eliminated.

④ The NTF must guarantee the stability of the 1-bit ΔΣ-modulator, that is the quantization error $e$ must be within the range of -1 to 0.

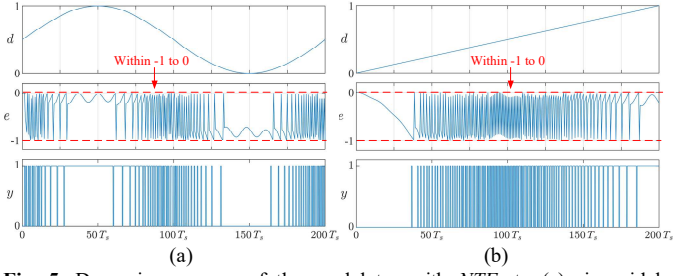

**Fig. 5.** Dynamic response of the modulator with $NTF_3$ to (a) sinusoidal excitations and (b) ramp excitations, where $\omega_e = 0.075$ for the case of $k = 0.15$.

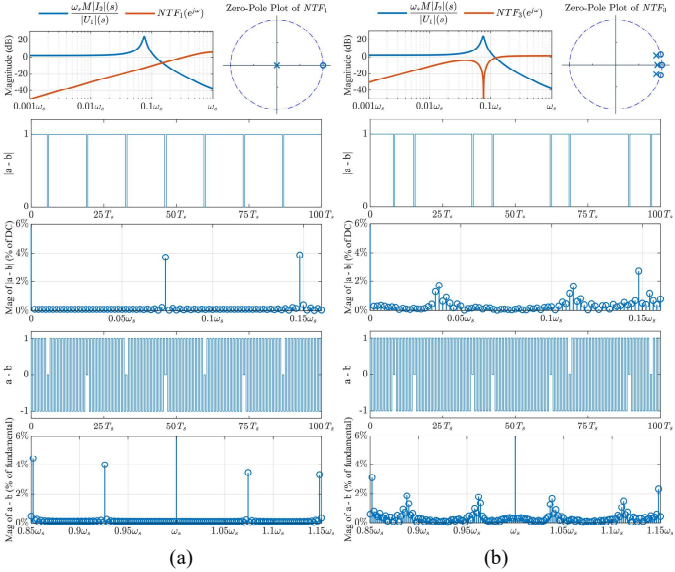

**Fig. 6.** Comparison of NTFs: bode plots, zero-pole diagrams, modulated waveforms and spectrums at $d = 0.963$: (a) $NTF_1$, and (b) $NTF_3$.

A third-order NTF satisfying the mentioned conditions is constructed as:

$$NTF_3(z) = \frac{z-1}{z-0.9} \frac{(z-e^{j\pi\frac{\omega_e}{\omega_s}})(z-e^{-j\pi\frac{\omega_e}{\omega_s}})}{(z-0.9e^{j\pi\frac{\omega_e}{\omega_s}})(z-0.9e^{-j\pi\frac{\omega_e}{\omega_s}})} \quad (8)$$

For the ΔΣ-modulator with 1-bit quantizer, the first-order modulator is unconditionally stable, while the bounds of stability of second-order modulator can be mathematically proven. But the conditions for stable operation of high-order NTFs are still unclear [6]. The stability of a high-order ΔΣ-modulator needs to be confirmed by extensive simulations.

For a completely resonant SS-compensated WPT system with $k = 0.15$, setting $\omega_e = 0.075\omega_s$ can suppress subharmonics that cause abnormal oscillations according to (2). In Fig. 5, the input pulse density $d$ varies as sinusoidal and ramp waveforms. The quantizer output $y$ tracks the input $d$, and the error $e$ always remains within the range of -1 to 0, which confirms the stability of the proposed NTF.

Fig. 6 compares the bode plots and pole-zero diagrams of $NTF_1$ and $NTF_3$, along with the waveforms and spectrums of modulated waves generated by the modulators applying $NTF_1$ and $NTF_3$. In the time domain, $NTF_3$ achieves noise shaping through distinct pulse arrangements, while maintaining a constant pulse density and a constant fundamental amplitude, which is guaranteed by the zero of $NTF_3$ at $z = 1$. With proper pole-zero placement, the $NTF_3$ achieves zero gain at $0.075\omega_s$, enabling the pulse density modulator to eliminate the frequency component near $0.075\omega_s$ in the modulated wave amplitude $|a - b|$. And the paired subharmonics near $0.925\omega_s$ and $1.075\omega_s$ in the spectrum of modulated wave $a - b$ are simultaneously canceled according to (1).

In summary, the proposed TSE-PDM method is based on a ΔΣ-modulator with a third-order NTF featuring notch characteristics instead of the conventional first-order NTF. The modulated wave $a - b$ contains no subharmonics near the resonant peaks of the resonant network's input admittance and output admittance, thus preventing abnormal oscillations.

## IV. Experimental Verification

A 220W SS-compensated WPT system was built for verifying the proposed method. The parameters of the prototype are listed in Table I.

The steady-state performances of the system under the conventional ΔΣ-PDM control and the proposed TSE-PDM control are tested at different operating points: $d_{1/2}$ ranging from 0.203 to 0.903 in steps of 0.02, and from 0.903 to 0.993 in steps of 0.01, while the pulse density of the other side was fixed at 1.

TABLE I
WPT SYSTEM PARAMETERS

| Symbol | Quantity | Value |
|---|---|---|
| $L_1, L_2$ | Resonant inductances | 31.7μH, 29.7μH |
| $C_1, C_2$ | Resonant capacitances | 8.88nF, 9.47nF |
| $R_1, R_2$ | Equivalent series resistance | 100mΩ |
| $V_g, V_o$ | DC side voltage | 50V |
| $k$ | Coupling coefficient | 0.15 |
| $f_s$ | Switching frequency | 300kHz |

### A. Steady-State Performances with Ideal Noise Shaping

Fig. 7 compares the fluctuations of resonant current amplitude under different pulse densities when applying conventional ΔΣ-PDM control and TSE-PDM control. The notch frequency of $NTF_3$ was precisely set at the resonant peak frequency $\omega_0$, i.e. $0.075\omega_s$. The proposed TSE-PDM method suppressed the fluctuations to within 22% over the entire test range of $d_{1/2}$, while the conventional ΔΣ-PDM could lead to a fluctuation of up to 60% in the worst case, i.e. at $d_{1/2} = 0.963$. Fig. 8 further compares the time-domain waveforms under the worst-case condition, demonstrating the superior performance of TSE-PDM.

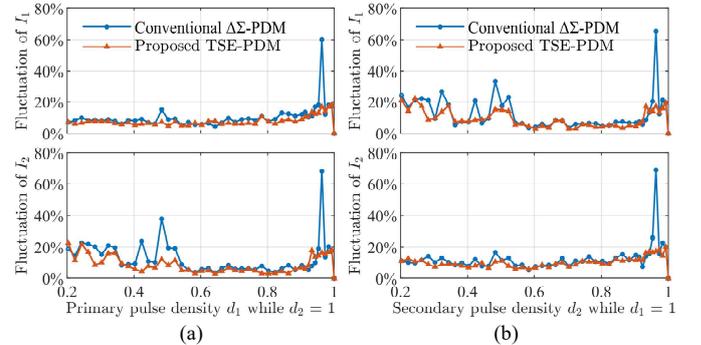

**Fig. 7.** The fluctuations in current amplitude of the conventional ΔΣ-PDM and TSE-PDM under (a) primary side control and (b) secondary side control.

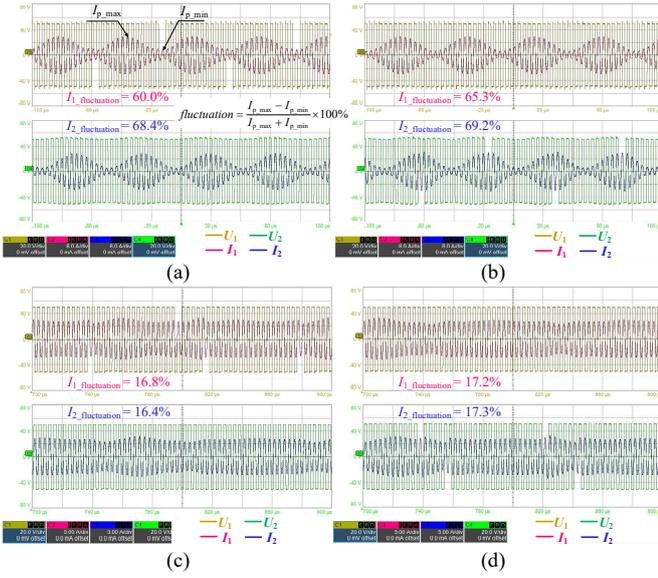

**Fig. 8.** Experimental waveforms under conventional ΔΣ-PDM control at (a) $d_1 = 0.963$ while $d_2 = 1$, and (b) $d_2 = 0.963$ while $d_1 = 1$. Experimental waveforms under proposed TSE-PDM control at (c) $d_1 = 0.963$ while $d_2 = 1$, and (d) $d_2 = 0.963$ while $d_1 = 1$ under conventional TSE-PDM control.

### B. Noise Shaping with Certain Deviations

The proposed TSE-PDM method exhibits a certain tolerance to the inaccurate identification of $k$, and overcomes the estimation error of $k$ in practical applications.

In Fig. 9, it was assumed that $k$ is inaccurately estimated as 0.13 and 0.17 (with a relative error of 13%), thus the notch frequency of $NTF_3$ was set to $0.065\omega_s$ and $0.085\omega_s$ according to (2). Compared to the ideal cases, the fluctuations of current amplitude were slightly increased, but still within 26% over the entire test range of $d_2$. In Fig. 10, the time-domain waveforms at $d_2 = 0.963$ is shown to demonstrate the effectiveness of TSE-PDM with deviations.

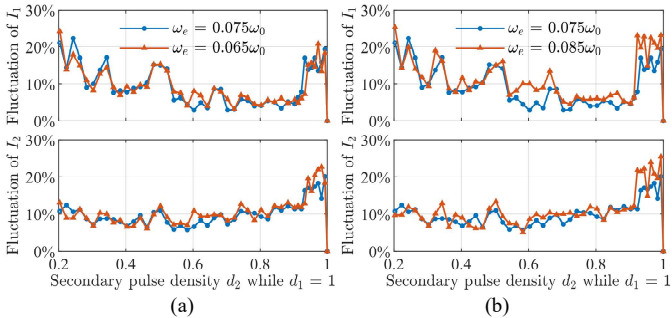

**Fig. 9.** Comparison of effects of TSE-PDM without and with deviations, where (a) $k$ is underestimated ($\omega_e = 0.065\omega_s$) and (b) $k$ is overestimated ($\omega_e = 0.085\omega_s$).

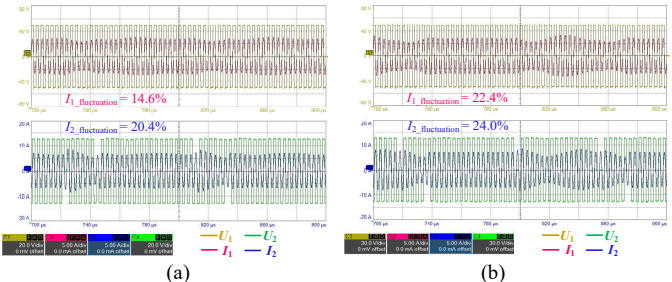

**Fig. 10.** Experimental waveforms at $d_2 = 0.963$ under TSE-PDM with deviations, where (a) $k$ is underestimated ($\omega_e = 0.065\omega_s$) and (b) $k$ is overestimated ($\omega_e = 0.085\omega_s$).

### C. Dynamic Responses

Fig. 11 shows the ΔΣ-PDM controlled WPT system's sinusoidal responses when $d_1 = 1$ and $d_2 = 0.5\sin(1000\pi t) + 0.5$ under the conditions of 15V input and output dc voltages. Compared with the conventional $NTF_1$-based system, the current of the TSE-PDM system with ideal noise shaping exhibits reduced oscillation, and the current envelope is basically consistent with $d_2$, demonstrating that the pulse density of modulated wave $U_2$ effectively tracks the variations in $d_2$. These characteristics show that the proposed TSE-PDM strategy exhibits wide modulation range and fast response.

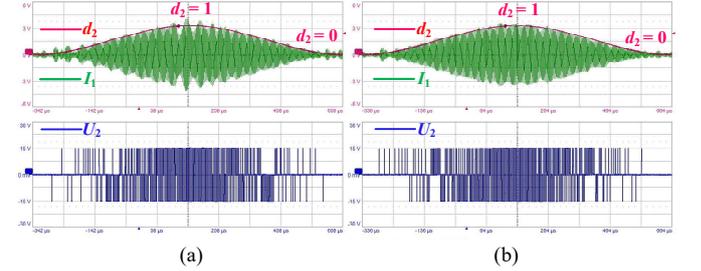

**Fig. 11.** Sinusoidal responses of (a) the conventional ΔΣ-PDM controlled system and (b) the TSE-PDM controlled system.

## V. CONCLUSION

This letter proposes a TSE-PDM method based on noise shaping for SS-compensated WPT systems. The subharmonic components that excite abnormal oscillations are eliminated through a third-order NTF featuring notch characteristics embedded in the ΔΣ-modulator. TSE-PDM is a simple and reliable method for preventing abnormal oscillations in resonant current, which exhibits fast dynamic response and certain tolerance to deviations caused by inaccurate coupling coefficient identification in NTF design.